\documentclass[pdflatex,sn-mathphys-ay]{sn-jnl}

\usepackage{graphicx}%
\usepackage{multirow}%
\usepackage{amsmath,amssymb,amsfonts}%
\usepackage{amsthm}%
\usepackage{makeidx}
\usepackage{mathrsfs}%
\usepackage[title]{appendix}%
\usepackage{xcolor}%
\usepackage{textcomp}%
\usepackage{manyfoot}%
\usepackage{booktabs}%
\usepackage{algorithm}%
\usepackage{algorithmicx}%
\usepackage{algpseudocode}%
\usepackage{listings}%
\usepackage{tabto}

\theoremstyle{thmstyleone}%
\theoremstyle{thmstyletwo}%

\theoremstyle{thmstylethree}%

\newcommand{\schrodinger}{Schr\"{o}dinger}

\newcommand{\GH}{\vspace{\baselineskip}\par G. H.  \tabto{20mm}}

\newcommand{\YA}{\vspace{\baselineskip}\par Y. A. \tabto{20mm}}

\makeindex
\begin{document}

\title[Theoretical Discovery, Experiment, and Controversy in the Aharonov-Bohm Effect]{Theoretical Discovery, Experiment, and Controversy in the Aharonov-Bohm Effect: \newline An Oral History Interview$^{\dagger}$ 
}

\author[1]{\fnm{Yakir} \sur{Aharonov}}

\author*[2]{\fnm{Guy} \sur{Hetzroni}}\email{guyhe@openu.ac.il}

\affil[1]{\orgdiv{Raymond and Beverly Sackler School of Physics
and Astronomy}, \orgname{Tel-Aviv University}, \orgaddress{\city{Tel-Aviv}, \postcode{69978}, \country{Israel}}}

\affil[2]{\orgdiv{Department of Natural Sciences}, \orgname{The Open University of Israel}, \orgaddress{\city{Ra'anana}, \postcode{4353701}, \country{Israel}}}

\abstract{This oral history interview provides Yakir Aharonov's perspective on the theoretical discovery of the Aharonov-Bohm effect in 1959, during  his PhD studies in Bristol with David Bohm, the reception of the effect, the efforts to test it empirically (up to Tonomura's experiment), and some of the debates regarding the existence of the effect and its interpretation.  The interview also discusses related later developments until the 1980s, including modular momentum and Berry's phase. It includes recollections from meetings with Werner Heisenberg, Richard Feynman, and Chen-Ning Yang, also mentioning John Bell, Robert Chambers, Werner Ehrenberg, Sir Charles Frank, Wendell Furry,  Gunnar K\"{a}ll\'{e}n, Maurice Pryce, Nathan Rosen, John Wheeler, and Eugene Wigner. }




\maketitle
{\let\thefootnote\relax\footnotetext{$^{\dagger}$This is a preprint of a paper forthcoming in The European Physical Journal H. Please cite the published version.}}

\section{Background: The Aharonov-Bohm effect}
In their 1959 paper, Yakir Aharonov and David Bohm introduced what is now known as the Aharonov–Bohm (AB) effect, a phenomenon wherein charged particles exhibit electromagnetic influence, manifested as a change in their quantum interference pattern, even in regions devoid of electric and magnetic fields. Their work has received extensive attention, with nearly 10,000 citations (as of April 2025). These citations reflect diverse research areas, including experimental verifications, applications in microscopy, foundational implications (on gauge theories, electromagnetism, quantum mechanics, theory of superconductivity), and philosophy of physics. The AB effect has also prompted debates over both its existence and interpretation. This interview\footnote{The interview  was conducted in a series of meetings in Tel-Aviv during January-March 2025. Small edits of language, structure, and clarity were made on Aharonov's request. } focuses on the period between mid-1950s and the 1980s, exploring Yakir Aharonov's perspective on the original discovery (Section 2), subsequent controversies (Sections 3, 5), early experimental realizations (Section 4), and its relation to some later developments in quantum foundations (Sections 6-7).

The AB effect fundamentally describes a phenomenon where electrons (or other charged particles) acquire a measurable phase shift while traversing paths in regions devoid of electromagnetic fields, yet enclosing a region in which an electric or magnetic field is present, as in a solenoid or a capacitor  (Fig. 1). This phase shift is proportional to the magnetic flux in the magnetic case, or, in the electric case, to the time integral of the electric potential difference.
\begin{figure}[h]
  \centering
  \includegraphics[width=\linewidth]{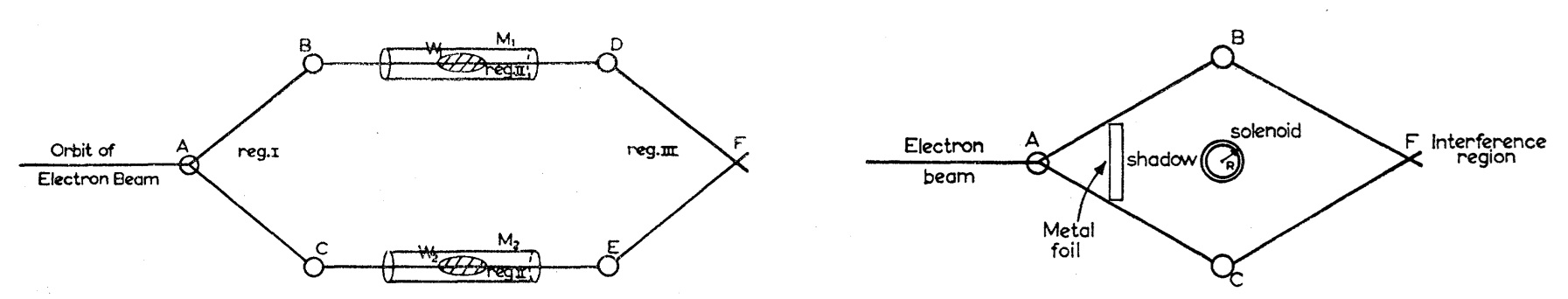}
  \caption{Left: The electric AB effect. Original caption: `Schematic experiment to demonstrate interference with time-dependent scalar potential. $A, B, C, D, E$: suitable devices to separate and divert beams. $W_1$, $W_2$: wave packets. $M_1, M_2$: cylindrical metal tubes. $F$: interference region.' Right: The magnetic AB effect. Original caption: `Schematic experiment to demonstrate interference with time-independent vector potential.' Reprinted with permission from \citet{aharonovbohm1959}, p. 486. Copyright (1959) by the American Physical Society.}
  \label{fig:ABeffect}
\end{figure}

Historically, the broad setting for the AB effect had been suggested in earlier, largely unnoticed works. As elucidated by \citet{hiley2013ABeffect}, the effect was initially hinted at in 1939 and the experiment that can measure the magnetic effect was explicitly described in a 1949 paper by \citet{ehrenberg1949refractive}. Yet, the effect remained largely unrecognized until Aharonov and Bohm discovered it, offered a thorough treatment, and highlighted its foundational significance.

\section{The discovery of the effect}
\label{discovery}

\GH You studied physics at the Technion from 1952 to 1956. Can we start with some background on how you came to study physics and work with David Bohm?

\YA I grew up in Kiryat Haim, and got interested in physics because I was interested in foundational questions, such as the meaning of life and free will. In high school, I came across \citeauthor{eddington1928natureHEB}'s (\citeyear{eddington1928natureHEB}) `The Nature of the Physical World' and `The Evolution of Physics' by \citet{einstein1939infeldHEB}. These books impressed me, especially Eddington. I  saw how physics helped him pose fundamental questions more clearly than the philosophers I had read before, and that was the reason I decided to study physics. 

I was an undergraduate student at the Technion in the first year of the science faculty, starting in ’52. In my third year I took a course in quantum theory taught by Nathan Rosen. At the end of the course, we had to write a short research paper. There were eight students then, and each one had to go to Nathan Rosen and get approval for the topic they wanted to write about. So when I came, I told him: `I want to do work on measurement problems in quantum theory'. He said: No, absolutely not. These are things that old people like me need to think about. You're young, do something real in physics... solid state physics... .' I told him: `Professor Rosen, I'm studying physics only because these questions interest me'. We reached an impasse; he wouldn't approve it. I remember walking down the corridor looking depressed. I met Gideon Carmi, who was then an assistant instructor in that course. He asked me what happened, I told him. He said: `You know what, I have an idea. Someone new just arrived at the Technion, named Bohm, go talk to him, maybe he will help you.'
So I went, knocked on the door, went in, introduced myself, and he asked me: `What interests you?' I started telling him what interested me. He said: `Okay, this interests me too. I'll talk to Rosen, I'll convince him to let you write this paper'. That’s how our  collaboration began. 

\GH In your time with Bohm at the Technion you did not yet work on the Aharonov-Bohm effect. 

\YA No. Our first paper was a commentary---a criticism of Dirac, who argued that in his equation the velocity of the electron is always $c$ or $-c$, because in the Hamiltonian there is $c\alpha \cdot p$. We said that that does not mean that it has a definite velocity $c$. 

Soon after, Bohm developed a very strong allergy to the sun and had to leave Israel. He got a position at Bristol University in England, and took me and Gideon Carmi with him to do the PhD there, and that's where Bohm and I started to develop the papers on EPR and the AB effect. 

\GH Your work on EPR \cite{aharonov1957epr,aharonov1960epr2} turned out to be very influential, we discuss it in a separate interview \citep{hetzroni2025aharonovquantum}. Can you now describe the path to what is now called the Aharonov-Bohm effect? 

\YA Already at the Technion, I was very excited about the work of Felix Bloch, who found out that when you have a periodic potential, there must be a gap in energy. This gap is the idea behind the explanation of all the beautiful phenomena of crystals, and it is the basic idea that led to transistors.

\GH This is something you learned in a solid state physics course in your undergrad at the Technion? 

\YA Yes. Possibly even from Rosen, or perhaps from Paul Zilsel, who was teaching solid state.

So anyhow, my thought was: what if instead of a  potential that is periodic as a function of position, we consider a potential that is  periodic as a function of the time variable? In this case you replace position by time, and thus you have to replace energy by momentum. When you have a periodic potential in position, you have a gap in the energy. I thought that maybe by looking at the time-dependent periodic potential that varies in  time instead of space, I would find that there must be gaps in momentum. By the way, that is true in the relativistic theory. But I looked at the solution of Schrödinger equation, added a periodic potential to the non-relativistic theory. Trying that,  I found that the only thing that happens is that the wave function is multiplied by a phase which is periodic in time. But a phase is not observable, since the observables are derived from terms such as $\psi^* \psi$, which do not depend on a time-dependent phase. Thus, a potential that was only a factor of time did not lead to anything interesting, and it really annoyed me.

And I remember one time I woke up in the morning---I often find many of my new ideas in my sleep--- and I thought of the possibility of having two different regions in space, such that in one region I have one time-dependent potential, and in another region I have a different time-dependent potential. Now I will have different phases, and then I could think about interference that will lead to an observable result. So I thought about the idea of having two Faraday cages. In a Faraday cage, when you put a charge inside, it is not affected at all by what's happening outside.

\GH Is this the origin of the first experiment described in the \cite{aharonovbohm1959} paper?

\YA At that time I thought about two real Faraday cages, each with a little door. You send an electron that can come through one of the doors or the other, or a superposition of them. You close the doors, so now the electron is inside, there is no electron outside. Thus, there’s only one electron existing simultaneously in both Faraday cages. We then introduce an electric field in the region between the two Faraday cages, and that will introduce a time-dependent potential inside the Faraday cages. Then you remove that field, so everything returns to free evolution. Then you open the little doors, the electron comes out, and lo and behold, when it interferes, the interference pattern indicates that it remembers what was the electric field at the time when it was not there. This was the beginning of the idea. 

I remember I came to Bohm---to David Bohm--- and told him about this idea. He said, ``Oh, that's really very interesting''. He came up with a modification of it, instead of having two Faraday cages, having two long cavities (see Fig. \ref{fig:ABeffect}, left). You send electrons through them as wave packets, and while they travel inside, you switch the potential difference and then switch it back off before the electrons come out. It’s closer to what you can really do in experiments.

\GH Basil \citet{hiley2013ABeffect} has a historical paper on the early history of the Aharonov-Bohm effect. He is writing that he heard from Lev Vaidman who heard it from you that at the time, you had spotted the potential producing observable effects, but that you did not realize that potentials were universally considered as mere mathematical artifacts. Would you say it is a good description?

\YA Yes, I had absolutely no idea about gauge transformations, or about the potentials being not physical, and that was actually the reason I discovered the effect. I could do it only because I did not know that time-dependent potentials are tricky, because they are just pure gauge. If I had known about the gauge, then I would never have thought about it. 

\GH You mean that the contribution of your ignorance about gauge freedom was that you had to think about spatial separation?

\YA  Yes, exactly.

\GH So when did you start thinking about the magnetic field?

\YA At the end of that year---1957 or 1958--- I went to a summer school in Oxford. That was a terrible time in physics; everybody was involved in what was called `dispersion relations'. At that time, there was a crisis in physics because any attempt to try and quantize the electromagnetic field led to infinities. So Heisenberg came with the idea that perhaps, in physics, you should not consider what happens at any instant of time. In such experiments, you always have the same setting: particles collide, and you see what happens after the collision. That’s the only experiment you could do in high-energy physics. So he invented the idea of what is called the S-matrix approach. People developed all kinds of mathematical theorems about what happens in scattering. And you need to use complex mathematics, and in many dimensions, it became terrible. All of physics at that time was just discussing more and more complicated mathematics. Nevertheless, Bohm suggested that I should go to this summer school to see what people were doing, to see if there was anything interesting. By the way,  I remember meeting John Bell there, and we had some discussion about my work with Bohm on the EPR paradox \citep{aharonov1957epr}, that was possibly the seed that turned later into his famous article \citep{bell1964}. But anyhow, at that school they discussed not only the scalar potential, but also the magnetic vector potentials. So when I heard about this, it occurred to me that maybe I could find an analogy to my two-cage setup but with a vector potential. That would be the magnetic situation where again the particle doesn’t move through the field. I came back to Bohm and told him about it---about the idea of having a magnetic field inside but only a vector potential outside. That made him very excited. He said, ``You know, we can now think about writing an article about it because it becomes really interesting.’’ 

\GH Do you remember anything else about the school in Oxford? Who was there? Did they also discuss nuclear interactions or only electrodynamics?

\YA I remember nothing at all. It was so boring... It was all about S-matrix theory. I think it was all in the context of electrodynamics. 

To make the article complete, Bohm suggested that I try to solve a scattering problem, in which we have a very, very thin solenoid, essentially a singular flux line, and an electron, initially described by a plain wave, is scattered by the vector potential generated by the flux line.

That turned out to be a very difficult problem to solve, because I had to sum up a series of fractional Bessel functions. Usually, a wave gets decomposed into a superposition of Bessel functions of integer order. But in our case, the order of the Bessel functions can be fractional, depending on the flux.\footnote{The fractional Bessel equation come about from the \schrodinger\ equation in cylindrical coordinates, with a vector potential gauge-fixed to the tangential direction. See Eq. (2) and below it in \cite{aharonovbohm1959}.} I looked at the encyclopedia of Bessel functions, and the only thing I could find there are superpositions of integer functions, not of series of fractional Bessel functions. 

I remember that at that time I met the chairman of the physics department at Bristol, Maurice Pryce, and I told him my problem. He helped me solve it by finding a differential equation whose solution could help us solve it. I don’t remember all the details now, but it’s all in the paper. I also suggested to him to be a co-author of the article,  but he said that he did not contribute enough.

\GH The way I was taught the Aharonov-Bohm effect, you can calculate the phase difference by exploiting the very concept of gauge invariance. One gauge transformation can make the potential vanish on one branch of the trajectory, another gauge transformation would make the potential vanish on the other branch, and the difference between these two gauge transformations at the interference region gives you the Aharonov-Bohm phase shift.\footnote{See, e.g., in \citet{BFFGHMS2023gauge}, \S2.2.2.} Indeed, many presentations of the effect spell it out from the outset in terms of gauge \citep{peshkin1989tonomura}. Did you think about it that way? 

\YA Yes. Once I got to that stage I really started to understand the issue of gauge transformations, and I realized that the effect is gauge invariant. By the way, today I don’t think about it as an effect of the vector potential any more. I think about it as a non-local effect of the electromagnetic field.\footnote{See \S\ref{sixties}. } 

\GH But back then it was different, right? Already from the title of the article, `Significance of the electromagnetic potentials in quantum theory' it seems to be about the way quantum mechanics allows us to better understand what electromagnetism is about.

\YA Right. At that time, I didn’t yet have this idea of non-locality. So it looks as if we showed that indeed, particles move in regions where there are only potentials and not fields, and \schrodinger\ equation is a local equation, so it seems to  tell you that the wave function is locally affected by the potential. It was very natural to assume that in quantum mechanics potentials play a role. Although, of course, it cannot be completely local, because I can perform a gauge transformation that replaces the potential from one side to the other side, while the physical effects remain the same. So obviously, although we have local equations where the wave function is local and the potential is local, there is a hidden non-locality due to gauge invariance. But anyhow, at that time our idea was that we had discovered that potentials play a fundamental role.

\section{Early reception}
\label{reception}

\GH From the literature it seems that your paper immediately invoked controversy. What notable reactions did you get at the time? 

\YA Yes, this caused some excitement.  Many people, when they first read about it, thought something might be wrong. I remember that Maurice Pryce, in fact, visited the Niels Bohr Institute in Copenhagen. He came back and told me: ``you know,  Bohr doesn’t believe in the effect, because it is against classical correspondence''. 

\GH Bohr thought that the effect violates his correspondence principle? 

\YA Yes. Pryce told me that in Copenhagen people were discussing the effect, some of them accepted it, and some of them, including Bohr, were against it. I am not sure what exactly were Bohr's reasons. I think it had to do with the reaction of the electrons on the solenoid. The solenoid is a classical object, right? If the effect were correct, classical correspondence implies that something must also happen to the classical object when many electrons go around it. Thus, that would somehow effectively contradict a classical limit, since we know that it can never happen classically, that the charge influences the solenoid without touching it. Only a few months later, Niels Bohr's son, Aage, who also later won the Nobel Prize for nuclear physics, finally managed to convince his father that the effect was real, using the argument by \citet{furry1960ab}.

\GH Other notable responses? 

\YA Bohm got a cable from 
Feynman. Feynman knew Bohm from some time they spent together in Princeton. He sent Bohm a cable that said `How beautiful, how come I never thought about it myself?'. In his lecture on the effect he is similarly stating that it seems strange in retrospect that no one thought of discussing the experiment before \citep[vol. II, \S15-2]{feynman1964lectures}. I got to meet Feynman several years later (\S\ref{sixties}). 

I also remember, Wigner did some calculations. He checked in fact to see whether quantum electrodynamics…  whether the correction due to quantum field theory, could change anything. I heard it from some people from Princeton. They told me that he did the calculation, decided that it's irrelevant. So he did not publish.

There was also a well-known Swedish physicist at the time, Gunnar K\"{a}ll\'{e}n. I believe he was important in quantum field theory. Sadly, he died few years later in a plane accident. He argued that the Aharonov-Bohm effect was not interesting because if the wave function is analytic then it always has a tail inside the field region. His idea was that by looking at what happens to this tail, which depends on the field, you can determine the outcome. He came to Copenhagen and gave a talk about it. He showed that if the wave function is analytic, you have the two packets moving from both sides, but you have the tail going inside. The wave function extends slightly into the region with the field, so the effect isn’t truly non-local. I think it’s like saying you can tell what someone is thinking just by looking at their fingernail; it’s an absurd way to approach physics. 

\GH Did you address this objection in your papers?

\YA No. I heard it only later. 

\GH I believe that a very similar argument is presented in \citet{strocchi1973superselection}. They are making a similar claim about the possibility to explain the effect using a nonvanishing tail. They are using the hydrodynamic formulation of quantum theory by \cite{madelung1927hydrodynamic}.

\YA All these approaches are just ways to avoid the real issue of non-locality. In the hydrodynamic formulation you define gauge invariant probability density $\rho$ and probability current $\vec{j}$, and write the \schrodinger\ equation using these variables.\footnote{The definitions are $\rho\equiv\psi^*\psi$ and $\vec{j}\equiv i\psi^* (\nabla-\frac{e}{c}\vec{A})\psi$.} In that representation, the only things that appear are the electric and magnetic fields, the vector potential doesn’t appear at all. Recently  in \citet{aharonov2024shushi} we showed that if you take a $\rho$ that has two bumps with a very small overlap in between, then in the limit in which $\rho$ in the overlap region in the middle goes to zero, the equations become completely unstable. The non-locality is replaced here by instability, the initial $\rho$ and $j$ do not determine what will happen later.

\GH Speaking of the hydrodynamics formulation of quantum mechanics, Bohmian mechanics, Bohm's interpretation \citep{bohm1952interpretation,bohm1993undivided} is closely linked to it, $\rho$ is identified with an actual distribution. Do you know if he saw any connection between his interpretation and the Aharonov-Bohm effect?

\YA Bohm never tried to make a connection between his hidden variable theory and the AB effect. There was a kind of tacit understanding between me and Bohm, we discussed very little about his  hidden variable theory, because I didn’t like certain aspects of his approach. 

\GH You also presented and discussed the effect in Xavier University in Cincinnati \citep{podolsky1962conference}. What do you recall from the event? 

\YA It was in 1962, organized by Podolsky, who was a professor at Cincinnati. At that time... it was before \citet{bell1964}, so people did not think that EPR \citep{epr1935} is very important. He wanted to have a conference to show that this thing that he discovered with Einstein and Rosen was still important. Wigner was there, and Dirac... I also met Furry there. It was really an interesting conference. Everett was there in the audience. We didn't have many discussions about his approach, but I think it was mentioned there. 

\begin{figure}
    \centering
    \includegraphics[width=1.0\linewidth]{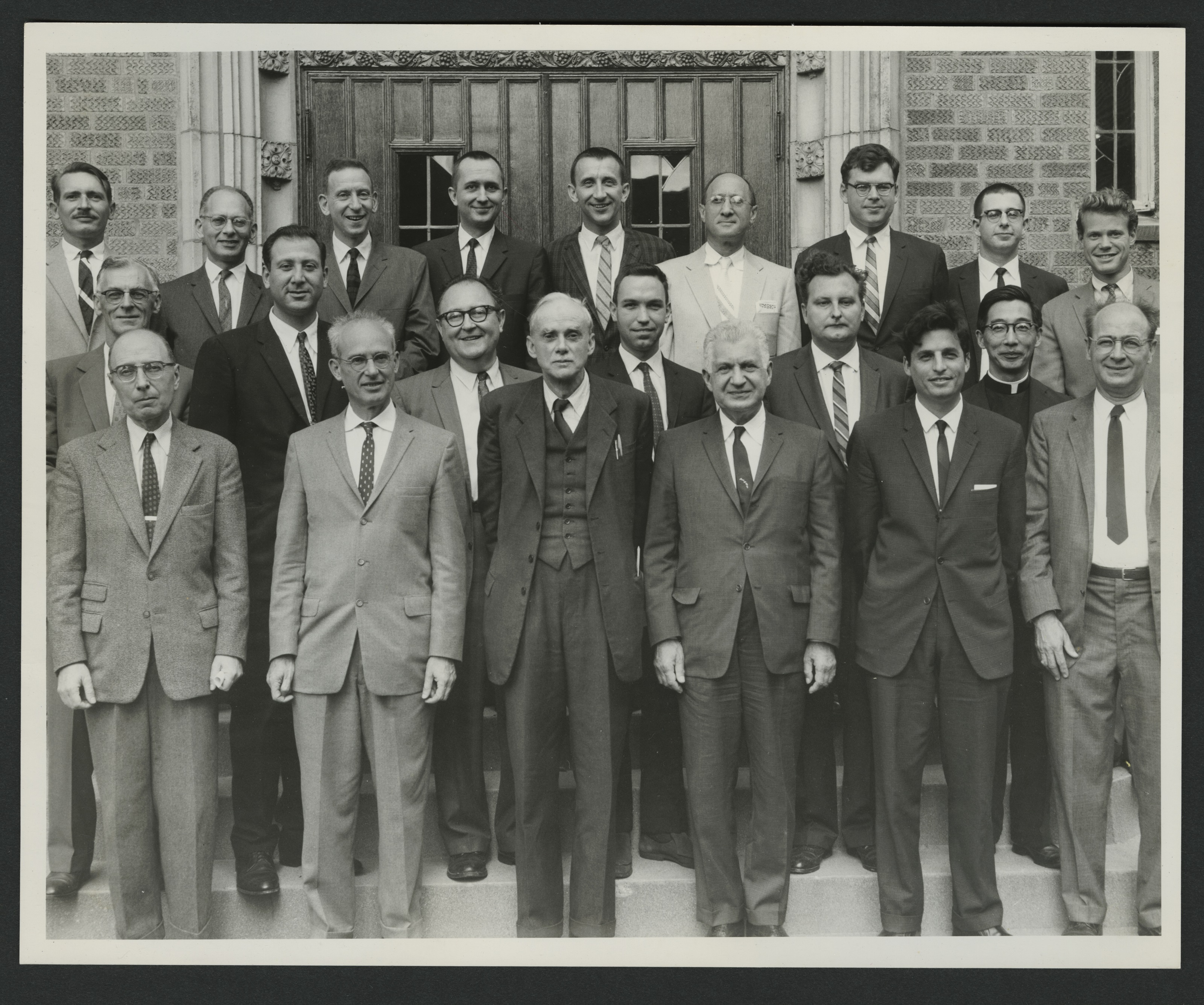}
    \caption{Xavier University, Cincinnati, 1962. Top Row (left-to-right): Jack Rivers, Harold Glaser, Eugen Merzbacher, John B. Hart, Jack A. Soules, Eugene Guth, Abner Shimony, Robert Podolsky, Austin Towle
Middle Row: William Band, Gideon Carmi, Solomon L. Schwebel, Dieter Brill, O.von Roos, Michael M. Yanase.
Bottom Row: 
Eugene P. Wigner, Nathan Rosen, P.A.M. Dirac, Boris Podolsky, Yakir Aharonov, Wendell H. Furry. 
Image and information courtesy of Xavier University Library \citep[see][]{podolsky1962conference}.  
}.
    \label{xavier1962}
\end{figure}

\YA At that meeting actually Podolsky was suggesting that instead of thinking about the local effect of the potential, you have to think about a non-local interaction between the magnetic field and the particle. 

\GH Did he see connection between the AB effect and EPR? 

\YA Yeah, he said that both of them are non-local in different ways. Podolsky was right, the EPR nonlocality does not change probabilities, and the Aharonov-Bohm nonlocality does change probabilities.  

\GH What else do you remember about the conference? 

\YA Dirac was very quiet. I did not talk to him at this conference. I remember that at that time I only talked with Wigner about my idea of a non-abelian gauge. I spoke with Furry as well, and Furry told me, `Oh, you should enjoy this day of yours, because it will never happen again... Every one of us gets one important idea in our life and that's it'. Gladly, he was completely wrong... 

\GH On a different matter, \citet{ehrenberg1949refractive} mentioned the magnetic effect at the end of their  paper, noting that it is curious that the equation `associates a phenomenon observable at least in principle with a flux', rather then with its temporal change. How and when did you come to know about their paper? 

\YA Sometimes after we published somebody wrote to Bohm about it. The minute we found about their paper we of course mentioned it in our later article \citet{aharonovbohm1961further-considerations}, and for a while Bohm was trying to say that we should call the effect Ehrenberg-Siday-Aharonov-Bohm effect. But then people said, well... The important thing was the significance that we saw in the effect, as a fundamental issue. There is also the electric effect, and also the exact solutions that I found. Because of all that, people decided that it should be called the Aharonov-Bohm effect. I also met Ehrenberg later in Birkbeck college (Siday has already passed away). I told him about the electric effect. He said `I don't believe it'. I believe it was obvious that they did not see the significance of the effect as a kind of fundamental problem of interaction. It was a small part of their article, and nobody paid attention to it. It's interesting, now, that there are many examples where people see something without realizing its significance.

\section{Measuring the effect}
\label{measuring}

\GH The first experimental tests of the effect came almost immediately after the publication of your paper. How did it happen? 

\YA We had at Bristol University meetings at four o'clock in the afternoon for tea, like in the old ages, where professors and graduate students were meeting together in the same room. In one meeting I told everybody about this idea. I said that I felt it will never be able to be tested experimentally, because if you have a solenoid, such that the electron can go around it either from the one side or from the other, the length difference between the paths will be too many wavelengths. I was afraid that nobody will be able to perform the experiment, because you could not have a solenoid that will be  thin enough to actually realize interference. 

\GH You thought that coherence would be lost in an actual experiment. 

\YA Exactly. But one professor there, Sir Charles Frank, an expert in solid state physics, said: `I know how you can do it. There is something called a ``magnetic whisker'' inside a crystal, a very very thin magnetic flux. And if you take one of those and set your electrons around it, then you will be able to do the experiment'. There was another professor sitting there, an experimentalist by name of Robert Chambers. He said, `I'm going to perform the experiment', and that's how it began.

While Chambers was conducting the experiment, I kept coming to his lab and making all kinds of suggestions... Until one day he told me `look, if you want me to ever finish the experiment, please go and continue with your theoretical work, and don't bother me with your experimental suggestions'. I admit, I was never particularly good at experiments. 

The results soon came to be as we predicted \citep{chambers1960ab}. Unfortunately, because it was finite, that whisker had leakage, so there was some magnetic field outside.

\GH Yes. Chambers mentions the possibility that the observed effect is due to leakage of the field. He attributes this possible objection to Pryce. Indeed, in the following years several experimental groups were trying to perform measurements with reduced leakage of the fields \citep{fowler1961ab,mollenstedt1962ABexperiment,bayh1962ab},  also with interference patterns in superconductors \citep{jaklevic1962ABinSC}.\footnote{See \citet{olariu1985fluxes} for a comprehensive review. } Were you engaged in these developments or in contact with these groups?

\YA I think I recall that two groups worked on it... One of them in Germany. They were able to actually build up a very narrow solenoid, sufficiently narrow that you could do the experiment around it and see the effect. It was not completely perfect; it was still finite, there was some magnetic leakage. Personally, I was sure that the effect is right. It was amusing and interesting to know that people are working on the experiment, but at this time I had already moved onto other theoretical developments and was not directly involved in the experiments. 

\GH A closely related issue was the measurements of the quantization of the magnetic flux, measured in superconducting rings \citep{deaver1961flux}. 

\YA Yes, Yang was able to explain this phenomenon using the Aharonov-Bohm effect \citep{yang1961byersFlux}. The idea is that if you have a charged particle going in the torus, and in the middle there is a magnetic flux---like half a fluxon for example---then you can show that the minimum energy is not zero, it corresponds to $\hbar/2$. Now imagine that in a superconductor you have a huge number of particles on the torus. And if there were non-quantized flux inside the hole, all of them would gain energy. So it is energetically preferable for them to create, in the inner surface of the superconductor, a small current that will complete the flux to an integer flux.\footnote{See \citep{aharonovrohrlich2008QP} \S4.6 for a pedagogical explanation.}  

\GH Many experimental papers from the 1960s highlighted leakage concerns, sought to minimize it and to measure its influence.

\YA Only Tonomura \citep{tonomura1986ab}  was able to arrange the experiment such that the flux was completely enclosed by a superconductor, and then there was no issue of leakage. 

\GH Could you describe your acquaintance with Akira Tonomura? Were you involved in his experimental work? 

\YA Tonomura was interested in the effect, not only for testing it, but also in using measurements of the AB phases for the purpose of microscopy,  to observe very small disturbances on the surface of magnetic fields.\footnote{On the later developments of this idea see \citet{tonomura2006ab-microscopy}.} He invited me to visit the Hitachi Laboratory in Japan. I visited. He later participated in the 50-year anniversary conference of the AB effect in Tel-Aviv.\footnote{A recording of Tonomura's talk at the event is found at: \url{https://video.tau.ac.il/events/index.php?option=com_k2&view=item&id=1329:the-ab-effect-and-its-expanding-applications}} We maintained our connection throughout the years, and I was deeply saddened when he developed cancer.

\section{Interpretation and Controversy}
\label{dispute}

\GH Okay, let's go back to the early 1960s. Following your original paper, you published three subsequent papers answering to various objections \citep{aharonovbohm1961further-considerations,aharonovbohm1962remarks,aharonovbohm1963further-discussion}. Most of the responses questioned your interpretation of the effect based on the potentials. Both Stanley \citet{mandelstam1962ab} and Bryce \citet{dewitt1962AB} offered interpretations of the effect in terms of quantum field operators, with a certain notion of nonlocality (with a subsequent discussion by \citealp{belinfante1962ab}).  In Weizmann Institute, \citet{peshkin1961ab} tried to explain the  effect based on a quantum description of the mechanical interaction  of the particle and the barrier or the source. 

However, beyond the interpretative debates, some of the papers \citep[including][]{furry1960ab,aharonovbohm1961further-considerations,peshkin1961ab} seem to try to push against a claim that I did not find in the literature from that time, that there will be no effect at all. You already mentioned (\S\ref{reception}) that Bohr held this position for a while. Where there other objections to the existence to the effect? 

\YA There were people who were trying to say the effect did not exist, because they claimed that the phase we calculated is cancelled by another phase. So there were people who argued that the effect isn’t real. Then there were others discussing whether it shows that the potentials are relevant or not. But most objections were, actually, that the effect is simply not consistent. 

\GH Can you elaborate on these objections? 

\YA I think the main objection at the time was that we only calculated the effect of  the source on the electron and did not calculate the backreaction, the effect of the electron on the source. People argued, `you're considering only the phase acquired by the electron as it goes past the capacitor, but you're ignoring the phase gained by the capacitor due to the electron’s influence'. People thought that maybe the two phases will cancel each other and we will not see anything. But it turns out that there are three phases, not two, when calculated correctly.\footnote{For example, in the case of the electric effect induced by a parallel plate capacitor, the two phases caused by the influence of the electron on each of the two sheets will cancel out. The overall phase is the one acquired by the electron that is familiar from the standard calculation \citep[\S3]{aharonovbohm1961further-considerations}.} 

\GH Who made this objection? 

\YA I think it was from a quite famous physicist at Bell labs. He wrote to us the objection. But later he said he discovered it was wrong.  The argument by \citet{furry1960ab} also convinced people that the effect is essential for consistency. 

\GH Okay... Do you want to say something also about the correspondence with \citet{peshkin1961ab}? I think they seemed to try to explain the effect somehow by a mechanical influence of the barrier between the electron and the solenoid, but they were also talking about the role of topology.

\YA Yes... They replaced the solenoid with a cylinder having charge distributed over its surface, so that rotating it produced a magnetic field. With two cylinders having opposite charges and rotating in opposite directions, it was possible to generate a magnetic flux inside, with no  field or forces acting outside. Then they have shown that you can understand the effect by looking at what the electron does to the cylinder. That was an argument essentially the same as the \citet{furry1960ab} gave for the electric effect, this time it was for the magnetic case. 

\GH People continued to object to the existence of the effect all the way until \citeauthor{tonomura1986ab}'s (\citeyear{tonomura1986ab}) experiment. For example \citet{bocchieri1984qlaws} argued that the existence of the effect contradicts the laws of quantum mechanics, and tried to explain existing experimental results based on the action of the Lorentz force on the part of the wavefunction penetrating the magnetic field \citep{bocchieri1982tonomura}.  

\YA I currently do not remember much about this debate... This was not the only objection at the time, but I replied to this one because they were very persistent but completely wrong. We showed this in a reply paper \citep{aharonov1984bocchieriReply}, but now I don't remember the details...

Generally, many people thought it very strange that particles could be affected non-locally by electric and magnetic fields. Historically, Newton thought that the sun can have a causal but nonlocal action on the earth, due to gravity.  However, following Maxwell's theory and special relativity, the accepted view became that in order not to violate causality you must have only local interactions with the field. The Aharonov-Bohm effect showed that due to the uncertainty principle, it is possible not to violate causality and still have nonlocal interactions. So, it was sort of going back to the idea of nonlocality that doesn't violate causality. This kind of quantum  nonlocal causality can exist in virtue of the thing that is effected nonlocally being completely uncertain.\footnote{For a comprehensive discussion see  \citet{aharonovrohrlich2008QP}, Chapter 6.} It violated the intuition of many people. 

\section{1960s: Feynman, gauge groups, modular variables, and Heisenberg}
\label{sixties}

\GH You mentioned before a meeting with 
Feynman. When did you meet him? 

\YA In 1964, I believe. I was invited to Los Angeles by some nuclear energy company that wanted  to set up a large group of theoreticians. They offered me four times what I had earned in Yeshiva University, but I did not take it because I was worried that I might not be free to work on my research.  But anyhow, at that time I visited Los Angeles, and I think I called Feynman. He was at that time at Pasadena, I believe. So I came to his office one afternoon, and we sat for a few hours. He asked me what I was working on and I told him that I like to invent quantum paradoxes. So he told me to tell him one. So I started to tell him, and he immediately stopped me and continued himself, describing what the paradox was. He was right. And then he said, `okay, I will answer your paradox'. He started to answer. At that time I knew that the answer was wrong. So I stopped him... We had this sort of `ping-pong' a number of times and I was very happy to feel that I was on a par with him. 

\GH What paradox did you discuss? 

\YA One was... Suppose to have a flux inside a superconductor, just like eventually Tonomura did in the experiment... in such a way that if the electron  moves outside, its electric field cannot penetrate into the solenoid, so it doesn't affect the solenoid at all. On the other hand, the solenoid does affect the electron, because the vector potential can penetrate. 

Now suppose we have a situation where inside the superconductor you start with a superposition of two fluxes. From the point of view of the electron, its phase shift measures whether it is an integer or a half-integer flux, so  making measurements outside does affect the solenoid, even though it is not allowed to, because the field cannot penetrate into the superconductor.\footnote{A detailed, yet more preliminary, discussion of this paradox appears in the recording of Aharonov's talk available in \citet{podolsky1962conference}. } 

It turns out that the solution to the paradox was that there is no way to have a superposition of integer and a half integer inside the superconductor. There is no meaning to it, it's a mixed state. 

\GH So, were you thinking at that period about other variations and extensions of the effect? 

\YA Yes, one important issue was the extension of the effect to non-Abelian gauge groups. I developed it with Daniel Wisnivesky, who was my first PhD student in Yeshiva University. In that paper \citep{aharonov1967wisnivesky}, we extended the AB effect to include non-Abelian gauge group, as in Yang-Mills theories \citep{yangmills1954}. 

\GH The paper also touches upon something that comes up a lot in the debates about the effect, which is the question of whether the effect is genuinely quantum. Often people say that the effect is in fact completely classical. This point of view is very different from the views presented in your original paper with Bohm and also in much later papers. But your paper with Wisnivesky does present the effect on a par with a classical analogue. Was your view on the matter different at the time? 

\YA The paper shows that there is something similar in classical physics. Due to general relativity, the frequency of a classical clock depends on the gravitational potential. The classical analogue is therefore based on two identical clocks that move in different trajectories and show different time later. Today it appears to me less similar than it did back then. The difference is that unlike in quantum theory, in this case, the change in the frequency is not a change in the energy. Also, quantum interference has no classical analogue, the classical example is not an interaction that changes probabilities.  The most important aspect of the AB effect is the change of dynamical variables in the nonlocal interaction. So even at that time, when I thought about the effect as an effect caused by the potentials, I did not think that the effect is completely classical. I soon found out that the classical analogues don't have the dynamical consequences; there is no change in any dynamical observable. The modular momentum is the observable that has no classical analogue. 

\noindent\fbox{\parbox{\textwidth 
}{
\footnotesize
\textbf{Box 1. Modular momentum and modular variables in quantum theory}

In the Heisenberg picture, instead of evolving the quantum state $|\psi\rangle$ itself, one describes the temporal evolution of a system by tracking the time-dependent operator $\mathcal O(t)$, of which the instantaneous state  is an eigenstate. Its dynamics follow the Heisenberg equation:
\begin{equation}
\label{HeisenbergEq}
\frac{d}{dt}\mathcal O(t) = \frac{i}{\hbar}\bigl[H, \mathcal O(t)\bigr].
\end{equation}

For a single non-relativistic particle, the elementary observables are the canonical coordinates $x$ and $p$, and any polynomial operator can be expressed as a sum of monomials $x^{m}p^{n}$ (or symmetrized variants thereof).

The double-slit interference experiment, where a wavefunction splits into two spatially disjoint packets $|\psi_L\rangle$ and $|\psi_R\rangle$, illustrates the limitations of these elementary operators. The expectation values of these operators for the coherent superposition state $|\psi_L\rangle+e^{i\alpha}|\psi_R\rangle$ do not depend on the non-local phase factor $\alpha$, although the resulting observable interference pattern does. To encode this non-local phase, one must include the translation operator:
\begin{equation}
T_L \equiv e^{i p L/\hbar},
\end{equation}

\begin{center}
\fbox{
\begin{minipage}[h]{0.7\textwidth}
    \includegraphics[width=\linewidth]{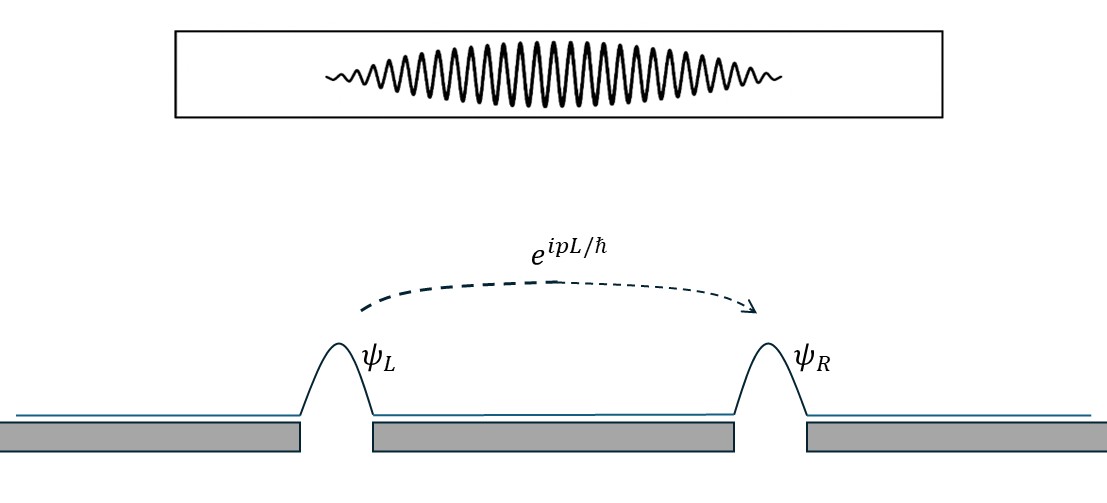}
    \centering
    \footnotesize 
\end{minipage}
}
\end{center}

\vspace{2mm}
Acting on a wavefunction, $T_L$ shifts its argument by $L$, mapping one wave-packet onto the other. The spectra of such operators lie on a circle, and their expectation values explicitly depend on $\alpha$, even when the packets remain separated. Because $e^{i p L/\hbar}$ is $2\pi$‑periodic in $pL/\hbar$, only the remainder of $p$ modulo $h/L$ is relevant:

\begin{equation}
p_{\mathrm{mod}} \equiv p \bmod \frac{h}{L},\qquad 0\le p_{\mathrm{mod}}<\frac{h}{L}.
\end{equation}

Thus, the pair $(p_{\mathrm{mod}}, N=\lfloor pL/h\rfloor)$ can replace the unbounded momentum variable. For a two-slit superposition $|\psi_L\rangle+e^{i\alpha}|\psi_R\rangle$ we have:
\begin{equation}
\langle e^{i p L/\hbar}\rangle = \frac{1}{2} e^{i\alpha}.
\end{equation}
Hence, $p_{\mathrm{mod}}$ is the minimal operator describing the non-local phase factor $\alpha$.

To demonstrate the dynamics consider, for example, a particle of charge~$e$ subject to a static electric potential~$V(x)$. The dynamics according to Eq. \eqref{HeisenbergEq} is
\begin{equation}
\frac{d}{dt}e^{i p L/\hbar}=\frac{i}{\hbar}\bigl[H,e^{i p L/\hbar}\bigr]
= \frac{i}{\hbar}\bigl[eV(x)-eV(x+L)\bigr]e^{i p L/\hbar} .
\end{equation}
Thus, even if the electric field vanishes along both paths, a non-uniform potential alters $p_{\mathrm{mod}}$ \emph{non-locally}, enabling a measurable phase shift. This is the electric Aharonov–Bohm effect.

Other modular variables include modular position $x_{\mathrm{mod}} \equiv x \bmod d$ (characterizing diffraction from gratings with spacing $d$), modular energy $E_{\mathrm{mod}} \equiv E \bmod \hbar\Omega$ (for systems driven by periodic fields of frequency $\Omega$), and modular velocity $v_{\mathrm{mod}}\equiv[(p-\frac{e}{c}A)/m]_{\mathrm{mod}}$ (crucial in magnetic vector potentials and encoding the gauge-invariant momentum for the magnetic Aharonov–Bohm effect; note that here $p-\frac{e}{c}A$ is taken modulo $h/L$).

Modular variables provide a natural framework for describing interference, explicitly exposing gauge-invariant phase information missed by ordinary $x$ and $p$ moments. Their non-local dynamics underlie the Aharonov–Bohm effects, offering clearer, operator-centric intuition into quantum interference. \citep[See][chapter 5.]{aharonovrohrlich2008QP}
}}

\GH You introduced the concept of modular momentum shortly after (see Box 1), and it was later published in \citet{aharonov1969modular}. Can you describe how you came to work with Pendleton and Petersen and developed this idea? 

\YA I met Aage Petersen already in 1960 when Bohm was invited to Copenhagen and he took me along. Petersen was an assistant of Bohr. I met Hugh Pendleton about one year later, after my PhD. I was a research associate at Brandeis University for one year, and Pendleton was a young professor there. Then, a few years later, I came again to Brandeis for a sabbatical, to spend one semester there. Peterson came to the United States to work with me, on a grant from Denmark. I think the work I did and when I had the idea of modular momentum was done during this semester I spent at Brandeis, and Peterson was there too. 

During that time I realised that while the \schrodinger\ picture is a very nice way of understanding the quantum evolution intuitively and mathematically, and it is  practical for solving problems, it is in fact the observables’ evolution that truly matters. There is indeed another way to describe a system: by asking `which observable the vector is an eigenstate of?' For example, you can find an observable---a function of position and momentum---such that when it acts on the vector, it yields a number, like a projection. This way it's possible to forget about the wavefunction and just follow that observable, in Heisenberg's picture. 

Then I thought on the situation of  two wave packets that are separated by distance $L$, as in the two parts of the wave that come through the two slits. In this case there is a difference between `one side plus the other' or `one side minus the other'.\footnote{Namely, the difference between the wave function $\psi_1+\psi_2$ and $\psi_1-\psi_2$, where the two terms are the spatially separated wavepackets.}  I started thinking about the question of which observables are sensitive to this relative phase. Notably, if the wavefunction completely vanishes between the two packets, then none of the usual observables, those constructed from powers of momentum and position, can capture the difference. To fully describe the situation, the relevant observables must include periodic functions of momentum. The Hermitian modular momentum operator $\cos (pL/\hbar)$ is such an observable. That’s how I found the significance of this new concept. There is a way to follow the evolution of the system by following the temporal evolution of this observable.

\GH Basically, putting all of the nonlocality into the operators. 

\YA Exactly. I replaced the whole idea of the wavefunction with the averages of this new modular momentum and its non-local equation. In the paper we argue that this is also the best way to describe what happens in the interaction between the electron and the solenoid in the AB effect. There is an exchange of modular momentum, and in fact, we know exactly when it happens. I still think so today. Look at the average of the modular velocity, which is defined using $[p-\frac{e}{c}A]_{\mathrm{mod}}$ (Box 1). It is gauge invariant, and when the two wavepackets move, and the line connecting them crosses the solenoid, there is a sudden change in the probabilities for this modular velocity. Of course, you can observe it only later. This description is in terms of nonlocal interaction of a gauge invariant quantity. 

I had, earlier at that time, the idea that you could explain what happens in interference, not by using the wave picture, but by saying the particle actually moves through one slit, but it `knows' whether the other slit is open or not, because it has also another observable, that will become measurable only later for the interference, but is already affected non-locally by whether the other slit is open or not.

\GH Maybe one could argue that this is sort of a Heisenberg-ian version of  Bohm's theory, definite position that is guided not by the wavefunction, but instead, by the nonlocal operators. 

\YA 
This is something that I kept thinking about over the years. If the electron is a wave smeared over different locations, is its charge also smeared? If so, the collapse of the wave function to a definite position would have resulted in electromagnetic radiation, so it makes no sense. This the important reason in my opinion to prefer a well defined location. 

There is, however, a difference from Bohm's view. For example, when a hydrogen atom is in an eigenstate of energy, the position of the electron in Bohm's picture doesn't change, there is no motion, but in Heisenberg's equation of motion it does move. Here we say that there is a definite position, we don't know where it is, but we can calculate it's equations of motion and they are very different from Bohm's.  

There is also another difference related to my later work on weak measurements \citep[chapter 16]{aharonov1989weak, aharonovrohrlich2008QP}. The point is that in contrast to Bohm, my approach does not need to make an additional postulate that the particle in the double-slit experiment has a definite position in one of the slits, but rather derives it using weak measurements of the interference pattern together with post-selection on the slit (`which-path') measurement \citep{aharonov2017twoslits}.

\GH I understand you got to present your modular-momentum account  of the double-slit experiment to Heisenberg. 

\YA 
Yes. I wrote the article with Pendleton and Petersen in 1965. Few years later I came to visit the Max Planck Institute. I was invited by someone there. And while I was there, I met Hans-Peter D\"{u}rr the assistant of Heisenberg at that time. When he heard about my idea, he said, `I would make sure that you will meet Heisenberg'. So then, indeed, next morning, I came to Heisenberg's office, and we started to talk. And then I asked `Professor Heisenberg, do you know how to describe interference in the two-slit experiment in your language?' He said, `No, I don't know how to do it'. 

Historically, a few months after  Heisenberg presented his theory of matrices in 1925 \citep{heisenberg1925matrix},  \schrodinger\ came and everybody left Heisenberg's matrices and started to use \schrodinger's  language for two reasons: one, that it was much, much simpler mathematically to solve complicated problems by using Schrödinger's approach, but also because the experiment of Davisson and Germer that showed wave interference of electrons, and then obviously you had to use the wave picture, and not the way that Heisenberg was describing it, by  his equations of motion for momentum and position. Basic variables are still the old position, momentum, but there is no wave element.

So I showed Heisenberg my idea of modular momentum. He was so excited that from then on, every visitor that would come to Heisenberg---he was still alive for maybe six, seven years later--- he would take him to the blackboard and show him my ideas how to actually use Heisenberg's approach to describe and solve the interference.

Heisenberg was always claiming that the \schrodinger\ approach is not really quantum mechanics, it is too classical, and that the essence of the theory is better captured by his matrices. I agree with him. The real difference between classical and quantum mechanics is in the dynamics. It's the commutator replacing the Poisson brackets. It is not just a matter of adding small corrections in $\hbar$, because when the particle is in a superposition of two separate wave packets, the relevant variable is the modular momentum. Modular variables have non-local features, and are described by fundamentally different equations of motion in quantum mechanics. In the limit where $\hbar\rightarrow 0$, they diverge and become completely unobservable.

\GH Did you also discuss the Aharonov-Bohm effect with Heisenberg?

\YA No.

\section{Later developments and reflections}
\label{later}

\GH You mentioned before (\S\ref{measuring}) your visit to Hitachi. This was, I believe, in the First International Symposium on the Foundations of Quantum Mechanics in the Light of New Technology, in 1983 \citep{kamefuchi1983HitachiSymposium}.  What do you recall from this visit? 

\YA Tonomura and I did not discuss the detail of his experiment, but he was very interested in all my ideas about the AB effect, and I gave a talk when I came to this conference in Hitachi \citep{aharonov1984nonlocal}. 

\begin{figure}
    \centering
    \includegraphics[width=1\linewidth]{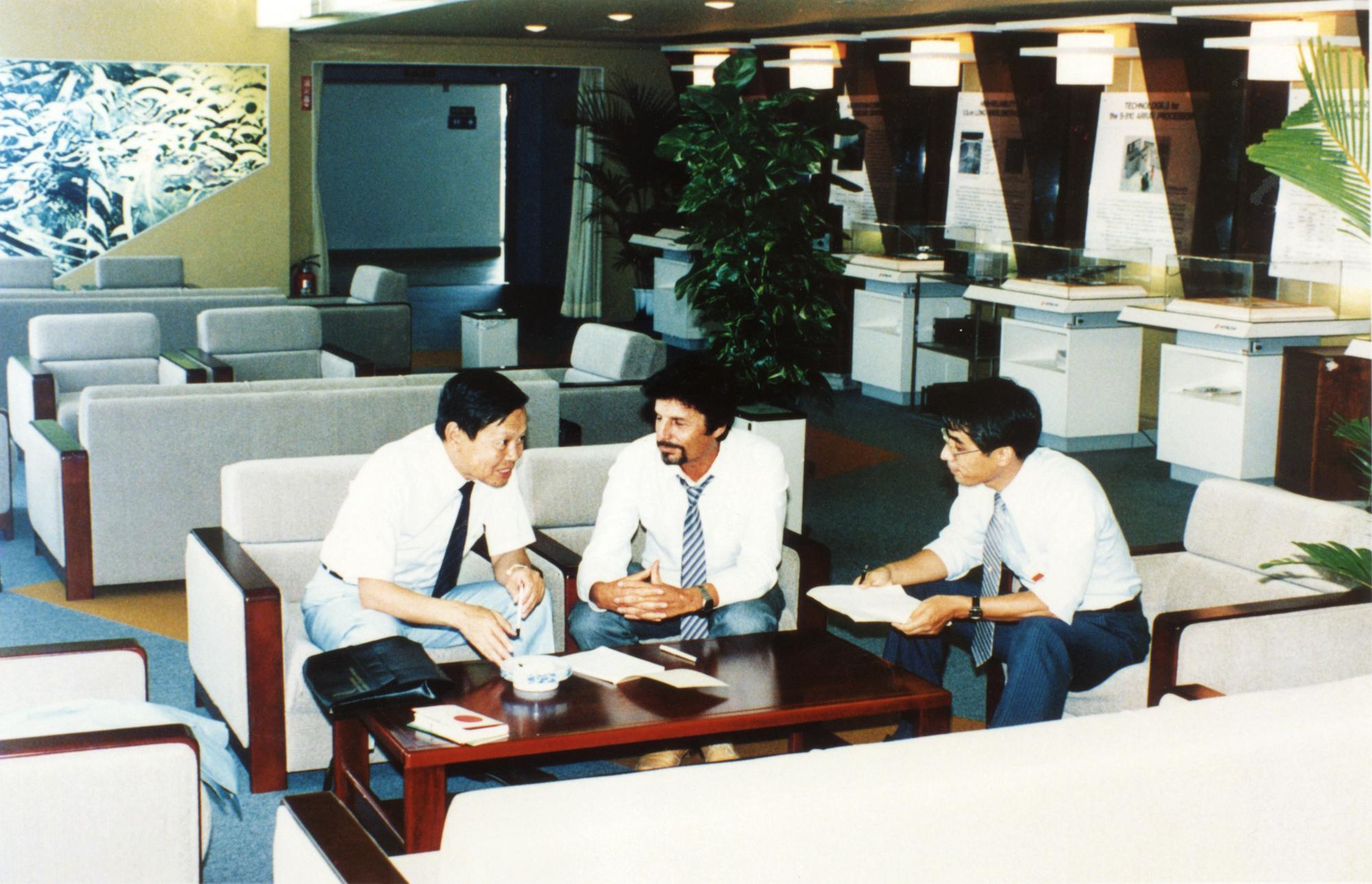}
    \caption{Left to right: Yang, Aharonov, and Tonomura in ISQM conference, Hitachi Central Research Laboratory, 1983. Courtesy of Hitachi, Ltd.; image originally appeard in \citet{fujikawa2014tonomura}. }
    \label{yang-aharonov-tonomura}
\end{figure}

That conference was also where I met Chen-Ning Yang. I remember we discussed the naming of the  effect. Yang was interested in the AB effect. He explained the quantization of the flux in a superconductor this way. He also wrote an article \citep{yang1970charge}, not knowing that I had done it before with my student  \citep{aharonov1967wisnivesky}, explaining the effect for non-abelian cases, like Yang-Mills theory \citep{yangmills1954}. During all these publications about the AB effect, he called it the ``Bohm-Aharonov effect''. When I met him in this conference in Hitachi I asked him, `professor Yang, why do you call it the Bohm-Aharonov, when the article is Aharonov-Bohm'? He said, `Oh, it’s Aharonov-Bohm alphabetically, but I'm sure the idea came from Bohm. He was your adviser'. I said, `No, it's not true. Actually, it was my idea'. He said, `Look, if you want me to believe it, tell Bohm to write me a letter, saying that actually it was your idea'. So I called Bohm, and Bohm sent him the letter, and from then on \citep[e.g.][]{wu2006interactions}, he started to call it `the Aharonov-Bohm effect', and he sent me a letter saying, `I apologize, from now on, I'll call it Aharonov-Bohm effect’.

\GH David Bohm, by the way,  says the same thing about the discovery of the effect in his interview to the American  Institute of Physics \citep{bohm1986interview}. But back to Yang... Do you remember discussing other topics with him, in Hitachi or elsewhere? 

\YA No, no. that was the only meeting we had. Yang expressed many times the view that the effect is important. Because physical interactions are gauge invariant, the effect is a general aspect of all interactions.

\GH What else do you recall from the Hitachi conference? 

\YA I remember meeting Wheeler for the first time. He was very excited, because he then just read about 
Berry's phase, and he told me: `Look, this is nearly as important as your effect'.  

I later met Wheeler a few more times, when he was in Texas and in Princeton. He was such a gentleman... When I was elected in 1993 to the National Academy of Science in United States, and it was announced, Wheeler sent me a letter saying that from now on it would be more interesting for him to come to the meeting of the National Academy.

\GH The paper on Berry's phase \citep{berry1984phase} was indeed published soon afterwards, and the Aharonov-Bohm effect is mentioned as a primary example. Yet, people seem to have different views on the relation between the two. 

\YA I think that the issues are very different. In the case of Berry's phase, he was relating to changes done adiabatically. The most interesting features of the AB effect, in contrast, is just when you do something very dramatically fast. You can see it through the modular momentum, that changes sharply in the case of the Aharonov-Bohm effect \citep{aharonov2004fluxline,aharonov2010nonlocalAB}, but gradually in the case of Berry's phase, where it does not play an important part. 

\GH Can you describe your acquaintance with Michael Berry? 

\YA We met for the first time much later, after I have already published several papers on Berry's phase. He invented the term ``super-oscillation" for my idea \citep{berry1992superoscillations}, and he wrote many articles about this with very sophisticated mathematics. The difference between us is that Michael loves to do all kinds of complicated mathematics, while I have a strong  physical intuition. We are, in some sense, complementary. Because of that, it was a very fruitful collaboration.  Later, Michael Berry and I got the 1995 EPS Europhysics Prize and the 1998 Wolf prize together, because somehow, the AB phase and Berry phase are considered to be related in some aspects.

\GH To conclude, would you like to say something on your current view on the significance of the effect?

\YA The effect teaches us that every fundamental interaction between two quantum systems always has two aspects. The first aspect is that of the local fields. The second is a purely quantum aspect, it is periodic and can thus be described by the modular variables. Consider for example the case in which an electron in the state of a plain  wave-function interacts with a non-isolated flux $\Phi$ with a circular cross-section of radius $R$.\footnote{See \citet{olariu1985fluxes} for a detailed treatment of the situation.} In this case the expectation value of the change in the electron momentum is proportional to the flux. This aspect can be understood classically. In principle, this aspect remains as the radius of the flux decreases (as long as the flux is kept constant): the interactions become rare, but the field, and thus the maximal momentum change, increases. However, this effect is not the whole story. In fact, this effect does not appear at all in the exact solution of \schrodinger\ equation in the case of $R=0$. In this case the cross section for interaction is periodic in the flux (proportional to $\sin^2{\frac{\pi e \Phi}{ch}}$).\footnote{See \citet{aharonovbohm1959}, Eq. 22.} This periodic aspect of the interaction changes the modular momentum, but not the expectation value of the momentum. Thus, a general interaction can be split into the AB effect, which is this bound and periodic influence of the infinitesimal string, and the classical local effect of the field. The sharper the interactions are---as a function of space and time---the quantum nonlocal aspect, becomes more significant, and the classical effect becomes more rare. It is quantum in the sense that in the classical limit this effect vanishes, as quantum coherence is lost. This can be seen for example from the divergence of the modular variables in the limit $\hbar\rightarrow 0$. The effect therefore reveals a basic aspect of all fundamental interactions, an aspect that was initially overlooked because it is counter-intuitive and only appears at the quantum level. We are used to thinking  about things classically.

\section*{Acknowledgements}
\label{acknowledgements}
GH would like to thank  Luisa Bonolis, Alex Blum, Giora Hon, Judy  Kupferman, and Bernadette Lessel. This research was supported by the Open University of Israel Research Fund (grant no. 515997) and by the ISF (grant no. 3445/24).

\makeatletter
\def\ps@plain{%
  \let\@oddhead\@empty
  \let\@evenhead\@empty
  \def\@oddfoot{\hfil\thepage\hfil}%
  \let\@evenfoot\@oddfoot}
\makeatother


\end{document}